\documentclass[preprint,12pt]{elsarticle}
\journal{Acta Materialia}

\usepackage{graphics,epsfig}
\usepackage{graphicx}
\usepackage{float}
\usepackage{amssymb,amstext,amsmath}
\usepackage{mathtools}
\usepackage{booktabs}
\usepackage{natbib}
\usepackage[latin1]{inputenc}
\usepackage{epstopdf}

\begin{document}

\begin{frontmatter}
\title{Thermodynamic dislocation theory: Size effect in torsion} 
\author{K.C. Le\,$^{a,b}$\footnote{Corresponding author. E-mail: lekhanhchau@tdt.edu.vn}, Y. Piao\,$^c$}
\address{$^a$\,Materials Mechanics Research Group, Ton Duc Thang University, Ho Chi Minh City, Vietnam\\
$^b$\,Faculty of Civil Engineering, Ton Duc Thang University, Ho Chi Minh City, Vietnam\\
$^c$\,Lehrstuhl f\"ur Mechanik - Materialtheorie, Ruhr-Universit\"at Bochum, D-44780 Bochum, Germany}

\begin{abstract}
The thermodynamic dislocation theory developed for non-uniform plastic deformations is used here for the analysis of twisted copper wires. With a small set of physical parameters that we expect to be independent of strain rate and temperature, we can simulate the torque-twist curves that match the experimental ones of Liu {\it et al.} (2012). It is shown that the size effect results from the accumulation and pile-up of excess dislocations.
\end{abstract}

\end{frontmatter}

\section{Introduction}
\label{Intro} 

The theory of dislocation mediated plasticity, originally proposed by Langer, Bouchbinder, and Lookman \cite{LBL-10} (called LBL-theory for short) and further developed in \cite{JSL-15}-\cite{Le18}, deals with the uniform plastic deformations of crystals driven by a constant strain rate. During these uniform plastic deformations the crystal may only have redundant dislocations with vanishing resultant Burgers vector. As shown in \cite{Le18a}-\cite{LPT18}, the extension of LBL-theory to more general thermodynamic dislocation theory (TDT) for non-uniform plastic deformations should account for excess dislocations due to the incompatibility of the plastic distortion \cite{Nye53}-\cite{Kroener55}. There are various examples of non-uniform plastic deformations in material science and engineering, the most typical of which being those in twisted wires \cite{Fleck94}-\cite{Liu2012} and bent beams \cite{Stoelken98}-\cite{Raabe2010}. Phenomenological approaches dealing with non-uniform plastic deformations include the strain gradient plasticity \cite{Fleck94,Fleck2001} (see also \cite{Aifantis1999}) which can capture the size effect. Its main weakness is that the chosen gradient of the plastic strain rate is not related to the dislocation density, so that the proposed constitutive equations are not based on the physics of dislocations. It is therefore impossible to derive the theory from the first principle calculation. The alternative continuum dislocation theory (CDT) accounting for excess dislocations, proposed for instance in \cite{Berdichevsky2006}-\cite{Liu2018}, is more predictive as kinematic hardening and size effect are captured by the first principle calculation of energy of dislocated crystals. However, since redundant dislocations and the disorder temperature are completely ignored, the above CDT cannot describe the isotropic hardening and the sensitivity of the torque-twist curves to temperature and strain rates. The purpose of this paper is to explore use of TDT for non-uniform plastic deformations \cite{Le18a}-\cite{LPT18} in modeling twisted copper wires. Our challenge is to simulate the torque-twist curves that show the hardening behavior and the size effect. We also want to compare these torque-twist curves with those obtained from simple extension of LBL-theory \cite{LPT18} and from the micro-torsion tests reported in \cite{Liu2012}. To make this comparison possible we will need to identify from the experimental data obtained in \cite{Liu2012} a list of material parameters for twisted copper wires.  For this purpose, we will use the large scale least-squares analysis described in \cite{Le17,Le18,LeTr17}. The comparison shows that: (i) the LBL-theory cannot predict the torque-twist curves and the size effect for wires in the micrometer range, (ii) the TDT provides an accurate prediction of the torque-twist curves and the size effect.

The thermodynamic dislocation theory is based on two unconventional ideas. The first of these is that, under nonequilibrium conditions, the atomically slow configurational degrees of freedom of dislocated crystals are characterized by an effective disorder temperature that differs from the ordinary kinetic-vibrational temperature. Both of these temperatures are thermodynamically well defined variables whose equations of motion determine the irreversible behaviors of these systems. The second principal idea is that entanglement of dislocations is the overwhelmingly dominant cause of resistance to deformation in crystals.  These two ideas have led to successfully predictive theories of strain hardening \cite{LBL-10,JSL-15}, steady-state stresses over exceedingly wide ranges of strain rates \cite{LBL-10}, thermal softening during deformation \cite{Le17}, yielding transitions between elastic and plastic responses  \cite{JSL-16,JSL-17}, shear banding instabilities \cite{JSL-17,Le18}, and size and Bauschinger effects \cite{Le18a}-\cite{LPT18}.  

We start in Sec.~\ref{EOM} with a brief annotated summary of the equations of motion to be used here.  Our focus is on the physical significance of the various parameters that occur in them.  We discuss which of these parameters are expected to be material-specific constants, independent of temperature and strain rate, and thus to be key ingredients of the theory. In Sec.~\ref{NI} we discretize the obtained system of governing equations and develop the numerical method for its solution. The parameter identification based on the large scale least squares analysis and the results of the numerical simulations are presented in Sec.~\ref{PI}.  We conclude in Sec.~\ref{CONCLUSIONS} with some remarks about the significance of these calculations.
 
\section{Equations of motion}
\label{EOM}
 
Suppose a thin polycrystalline copper wire with a circular cross section $A$, of radius $R$ and length $L$, is subjected to torsion (see the wire with its cross-section in Fig.~1). For this particular geometry of the wire and under the condition $R\ll L$ it is natural to assume that the warping of the wire vanishes, while the circumferential displacement is $u_\varphi=\omega rz$, with $\omega $ being the twist angle per unit length. Thus, the total shear strain of the wire $\gamma =2\epsilon_{\varphi z}=\omega r$ and the shear strain rate $\dot{\gamma}=\dot{\omega}r$ turn out to be non-uniform as they are  linear functions of radius $r$. 

\begin{figure}[htp]
	\centering
	\includegraphics[width=.5\textwidth]{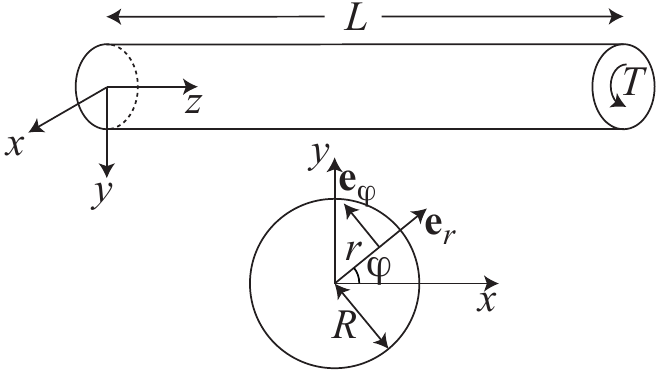}
	\caption{Torsion of a wire.}
	\label{bar}
\end{figure}

Now, let this system be driven at a constant twist rate $\dot\omega \equiv \phi_0/t_0$, where $t_0$ is a characteristic microscopic time scale. Since the system experiences a steady state torsional deformation, we can replace the time $t$ by the total twist angle (per unit length) $\omega$ so that $t_0\,\partial/\partial t \to \phi_0\,\partial/\partial \omega$. The equation of motion for the flow stress becomes
\begin{equation}
\label{tauydot}
\frac{\partial \tau_Y}{\partial \omega} = \mu\,\left[r - \frac{q(\gamma )}{\phi_0}\right],
\end{equation}
with $\mu$ being the shear modulus. This equation is derived from Eq. (II.1) in \cite{LeTr18} by replacing $\gamma =r \omega$ and multiplying both sides by $r$. Note that for uniform plastic deformations involving only redundant dislocations $q(\gamma )/t_0$ equals the plastic shear rate $\dot{\beta}$, with $\beta $ being the uniform plastic distortion. However, if $\beta $ is non-uniform, it is not necessarily so.

The state variables that describe this system are the elastic strain $\gamma-\beta$, the areal densities of redundant dislocations $\rho_r$ and excess dislocations $\rho_g\equiv  |\beta _{,r}+\beta/r|/ b$ (where $b$ is the length of the Burgers vector), and the effective disorder temperature $\chi$ (cf. \cite{JSL-16,Kroener1992}). All four quantities, $\gamma-\beta$, $\rho_r$, $\rho_g$, and $\chi$, are functions of $r$ and $\omega$. 

The central, dislocation-specific ingredient of this analysis is the thermally activated depinning formula for $q$ as a function of a flow stress $\tau_Y$ and a total dislocation density $\rho=\rho_r+\rho_g$:  
\begin{equation}
\begin{split}
\label{qdef}
q(\tau_Y,\rho)= b\sqrt{\rho} [f_P(\tau_Y,\rho)-f_P(-\tau_Y,\rho)], 
\\
f_P(\tau_Y,\rho)=\exp\,\Bigl[-\,\frac{1}{\theta}\,e^{-\tau_Y/\tau_T(\rho)}\Bigr]. 
\end{split}
\end{equation}
This is an Orowan relation of the form $q = \rho\,b\,v\,t_0$ in which the speed of the dislocations $v$ is given by the distance between them multiplied by the rate at which they are depinned from each other. That rate is approximated here by the activation terms $f_P(\tau_Y,\rho)$ and $-f_P(-\tau_Y,\rho)$, in which the energy barrier $e_P=k_BT_P$ (implicit in the scaling of $\theta=T/T_P$) is reduced by the stress dependent factor $e^{-\tau_Y/\tau_T(\rho)}$, where  $\tau_T(\rho)= \mu_T\,b \sqrt{\rho}$ is the Taylor stress with $\mu_T$ being proportional to $\mu $ (see Section \ref{NI}). Note that antisymmetry is required in Eq.~(\ref{qdef}), especially when dealing with the load reversal, both to preserve reflection symmetry, and to satisfy the second-law requirement that the energy dissipation rate, $\tau_Yq/(\phi_0r)$, is non-negative.  

The pinning energy $e_P$ is large, of the order of electron volts, so that $\theta$ is very small.  As a result, $q(\tau_Y,\rho)$ is an extremely rapidly varying function of $\tau_Y$ and $\theta$.  This strongly nonlinear behavior is the key to understanding yielding transitions and shear banding as well as many other important features of crystal plasticity.  For example, the extremely slow variation of the steady-state flow stress as a function of strain rate discussed in \cite{LBL-10} is the converse of the extremely rapid variation of $q$ as a function of $\tau_Y$ in Eq.~(\ref{qdef}).  

The equation of motion for the total dislocation density $\rho=\rho_r+\rho_g$ describes energy flow. It says that some fraction of the power delivered to the system by external driving is converted into the energy of dislocations, and that that energy is dissipated according to a detailed-balance analysis involving the effective temperature $\chi$. In terms of the twist angle $\omega$ this equation reads: 
\begin{equation}
\label{rhodot}
\frac{\partial \rho}{\partial \omega} = K_\rho \,\frac{\tau_Y\,q}{a^2\nu(\theta,\rho,\phi_0r)^2\,\mu\,\phi_0}\, \Bigl[1 -\frac{\rho}{\rho_{ss}(\chi)} \Bigr],
\end{equation}
with $\rho_{ss}(\chi) =(1/a^2)e^{- e_D/\chi}$ being the steady-state value of $\rho$ at given $\chi$, $e_D$ a characteristic formation energy for dislocations, and $a$ denoting the average spacing between dislocations in the limit of infinite  $\chi$ ($a$ is a length of the order of tens of atomic spacings). The coefficient $K_\rho $ is an energy conversion factor that, according to arguments presented in  \cite{LBL-10} and \cite{JSL-17}, should be independent of both strain rate and temperature. The other function that appears in the prefactor in Eq.~(\ref{rhodot}) is
\begin{equation}
\label{nudef}
\nu(\theta,\rho,q_0) \equiv \ln\Bigl(\frac{1}{\theta}\Bigr) - \ln\Bigl[\ln\Bigl(\frac{b\sqrt{\rho}}{q_0}\Bigr)\Bigr].
\end{equation}

The equation of motion for the effective temperature $\chi$ is a statement of the first law of thermodynamics for the configurational subsystem: 
\begin{equation}
\label{chidot}
\frac{\partial \chi }{\partial \omega} = K\,\frac{\tau_Y e_D \,q}{\mu\,\phi_0}\,\Bigl( 1 -\frac{\chi}{\chi_0} \Bigr). 
\end{equation}
Here, $\chi_0$ is the steady-state value of $\chi$ for strain rates appreciably smaller than inverse atomic relaxation times, i.e. much smaller than $t_0^{-1}$. The dimensionless factor $K$ is inversely proportional to the effective specific heat $c_{e\!f\!f}$. Since the maximum shear strain rate (reached at the outer radius of the bar) for the small twist rate in our torsion test is small, we assume that $K$ is a constant.  

The equation for the plastic distortion $\beta$ reads 
\begin{equation}
\label{microforces}
\tau -\tau_B -\tau_Y =0.
\end{equation} 
This equation is the balance of microforces acting on excess dislocations which, together with the boundary condition at the free boundary $r = R$, can be derived from the variational equation for irreversible processes \cite{Le18a,LP18}. Since dislocations can reach the free boundary and form an array of dislocations there, we add the surface energy to the energy functional (per unit length)
\begin{equation}
\label{energyfunc}
I=\int_A \psi (\gamma -\beta,\rho _r,\rho _g,\chi )\, da+\int_{\partial A} \Gamma (\beta)\, ds,
\end{equation}
where
\begin{equation}
\label{freeenergy}
\psi (\gamma -\beta,\rho _r,\rho _g,\chi )=\frac{1}{2}\mu (\gamma -\beta)^2+\gamma_D\rho_r+\psi_m(\rho_g)-\chi (-\rho \ln (a^2\rho )+\rho )/L
\end{equation}
is the bulk free energy density, with $\gamma _D = e_D/L$, and
\begin{equation}
\label{surfaceenergy}
\Gamma (\beta )=\frac{\mu b}{4\pi} \beta \left( \ln \frac{e\beta _*}{\beta }+\frac{1}{2}\alpha \beta \right)
\end{equation}
the modified Read-Shockley surface energy density of an array of screw dislocations (having the surface dislocation density $\beta /b$ and forming the twist boundary). The log-term in \eqref{surfaceenergy} was obtained by Read and Shockley \cite{Read1950}, while the $\beta ^2$-term by Vitek \cite{Vitek1987}. We also introduce the dissipation function
\begin{equation}
\label{dissipation}
D(\dot{\beta},\dot{\rho},\dot{\chi})=\tau_Y \dot{\beta}+\frac{1}{2}d_\rho \dot{\rho}^2+\frac{1}{2}d_\chi \dot{\chi}^2.
\end{equation}
We require that the above equations \eqref{rhodot}, \eqref{chidot}, and \eqref{microforces} obey the variational equation \cite{Le18a,LP18}
\begin{equation}
\label{vequation}
\delta I+\int_A \left( \frac{\partial D}{\partial \dot{\beta}}\delta \beta +\frac{\partial D}{\partial \dot{\rho}}\delta \rho + \frac{\partial D}{\partial \dot{\chi}}\delta \chi \right) \, da =0.
\end{equation}

Using the standard calculus of variations one can easily derive the above equations from \eqref{vequation} provided the coefficients $d_\rho$ and $d_\chi $ are appropriately chosen. In
Eq.~\eqref{microforces} obtained from the variation of $\beta $ the first term $\tau=\mu(\gamma -\beta)=\mu(\omega r-\beta)$ is the applied shear stress, the second term the back-stress due to the interaction of excess dislocations, and the last one the flow stress. The back stress is given by
\begin{equation}
\label{zeta}
\tau_B =-\frac{1}{b^2}\frac{\partial ^2\psi_m}{\partial (\rho_g)^2}(\beta_{,rr}+\beta_{,r}/r-\beta/r^2),
\end{equation}
with $\psi_m$ being the free energy density of excess dislocations. Note that the applied shear stress is equal to the flow stress for the uniform plastic deformations. Berdichevsky \cite{VB17} has found $\psi_m$ for the locally periodic arrangement of excess screw dislocations in a bar under torsion. However, as shown by us in \cite{LP18}, his expression must be extrapolated to the extremely small or large dislocation densities to guarantee the existence of solution within TDT. The extrapolated energy density of excess dislocations reads \cite{LP18}
\begin{equation}
\label{energydiss}
\psi_m(\rho_g)=\mu b^2 \rho_g \left( \phi_*+\frac{1}{4\pi }\ln \frac{1}{k_0+b^2\rho_g}\right) +\frac{1}{8\pi}\mu k_1(b^2\rho_g)^2,
\end{equation}
with $\phi_* =-0.105$, $k_0$ being a small constant correcting the behavior of the derivative of energy at $\rho_g = 0$, and $k_1$ another constant correcting the behavior of the energy at large densities of the excess dislocations. Using \eqref{energydiss} we find that $\tau_B$ is given by
\begin{equation}
\label{backstress}
\tau_B=-\mu b^2\frac{k_1\xi^2+(2k_0k_1-1)\xi+k_1k_0^2-2k_0}{4\pi(k_0+\xi)^2}(\beta_{,rr}+\beta_{,r}/r-\beta/r^2),
\end{equation}
where $\xi=b|\beta_{,r}+\beta/r|$. Equation (\ref{microforces}) is subjected to the boundary conditions $\beta(0)=0$ and 
\begin{equation}
\label{boundaryc}
\left. \left( \frac{\partial \psi_m}{\partial \rho_g}+b \frac{\partial \Gamma}{\partial \beta}\right) \right|_{r=R}=\gamma_D.
\end{equation}
Condition \eqref{boundaryc} is obtained from \eqref{vequation} as the natural boundary condition. With $\psi_m$ from \eqref{energydiss} and $\Gamma$ from \eqref{surfaceenergy} we get
\begin{equation}
\label{bound1}
\frac{\partial \psi_m}{\partial \rho_g}=\mu b^2\left[ -\frac{\xi}{4\pi(k_0+\xi)}-\frac{\ln (k_0+\xi)}{4\pi}+\frac{k_1\xi}{4\pi}-0.105\right] ,
\end{equation}
and
\begin{equation}
\label{bound2}
b\frac{\partial \Gamma}{\partial \beta}=\frac{1}{4\pi }\mu b^2 \left( \ln \frac{\beta _*}{\beta }+\alpha \beta \right) .
\end{equation}

In this paper we will consider also a simple extension of LBL-theory to non-uniform deformations proposed in \cite{LPT18}, where the excess dislocations are ignored. Since the back-stress $\tau_B=0$ in this case, we identify $\tau_Y$ with $\tau $ in Eqs.~\eqref{tauydot}-\eqref{chidot}. This system of equations becomes close, and after its integration we use $\tau_Y=\tau$ to compute the torque.

\section{Discretization and method of solution}
\label{NI}

For the purpose of numerical integration of the system of equations (\ref{tauydot})-(\ref{boundaryc}) let us introduce the following variables and quantities
\begin{equation}
\begin{split}
\tilde{r}=r/R,\quad \tilde{\tau}=\tau/\mu, \quad \tilde{\tau}_Y=\tau_Y/\mu ,\quad \tilde{\tau}_B=\tau_B/\mu , 
\\
\tilde{\omega}=R\omega , \quad \tilde{\chi}=\frac{\chi}{e_D}, \quad \eta=\frac{b}{R},\quad \tilde{\rho}=a^2\rho .\label{dimless}
\end{split}
\end{equation}
The variable $\tilde{r}$ changes from zero to 1. The  dimensionless quantity $\tilde{\omega }$ has the meaning of the maximum shear strain achieved at the outer radius. The calculation of the dimensionless torque $\tilde{T}=T/R^3$ as function of $\tilde{\omega}=\omega R$ is convenient for the later comparison with the experimental data from \cite{Liu2012}. Then we rewrite Eq.~(\ref{qdef}) in the form
\begin{equation}
\label{tildeq}
q(\tau_Y,\rho)=\frac{b}{a}\tilde{q}(\tilde{\tau}_Y,\tilde{\rho}),
\end{equation}
where
\begin{equation}
\label{tildeqdef}
\tilde{q}(\tilde{\tau}_Y,\tilde{\rho})=\sqrt{\tilde{\rho}}[\tilde{f}_P(\tilde{\tau}_Y,\tilde{\rho})-\tilde{f}_P(-\tilde{\tau}_Y,\tilde{\rho})].
\end{equation}
We set $\tilde{\mu}_T=(b/a)\mu_T=\mu s$ and assume that $s$ is independent of temperature and strain rate. Then
\begin{equation}
\label{tildefp}
\tilde{f}_P(\tilde{\tau}_Y,\tilde{\rho})=\exp\,\Bigl[-\,\frac{1}{\theta}\,e^{-\tilde{\tau}_Y/(s\sqrt{\tilde{\rho }})}\Bigr].
\end{equation}
We define $\tilde{\phi}_0=(a/b)R\phi_0$ so that $q/(R\phi_0)=\tilde{q}/\tilde{\phi}_0$. Function $\nu $ in Eq.~(\ref{nudef}) becomes
\begin{equation}
\label{nudef1}
\tilde{\nu}(\theta,\tilde{\rho},\tilde{\phi}_0\tilde{r}) \equiv \ln\Bigl(\frac{1}{\theta}\Bigr) - \ln\Bigl[\ln\Bigl(\frac{\sqrt{\tilde{\rho}}}{\tilde{\phi}_0\tilde{r}}\Bigr)\Bigr].
\end{equation}
The dimensionless steady-state quantities are
\begin{equation}
\label{ss}
\tilde{\rho}_{ss}(\tilde{\chi})=e^{-1/\tilde{\chi}}, \quad \tilde{\chi}_0=\chi_0/e_D.
\end{equation}
Using $\tilde{q}$ instead of $q$ as the dimensionless measure of plastic strain rate means that we are effectively rescaling $t_0$ by a factor $b/a$. For purposes of this analysis, we assume that $(a/b)t_0=10^{-12}$s.

\begin{figure}[htp]
	\centering
	\includegraphics[width=.65\textwidth]{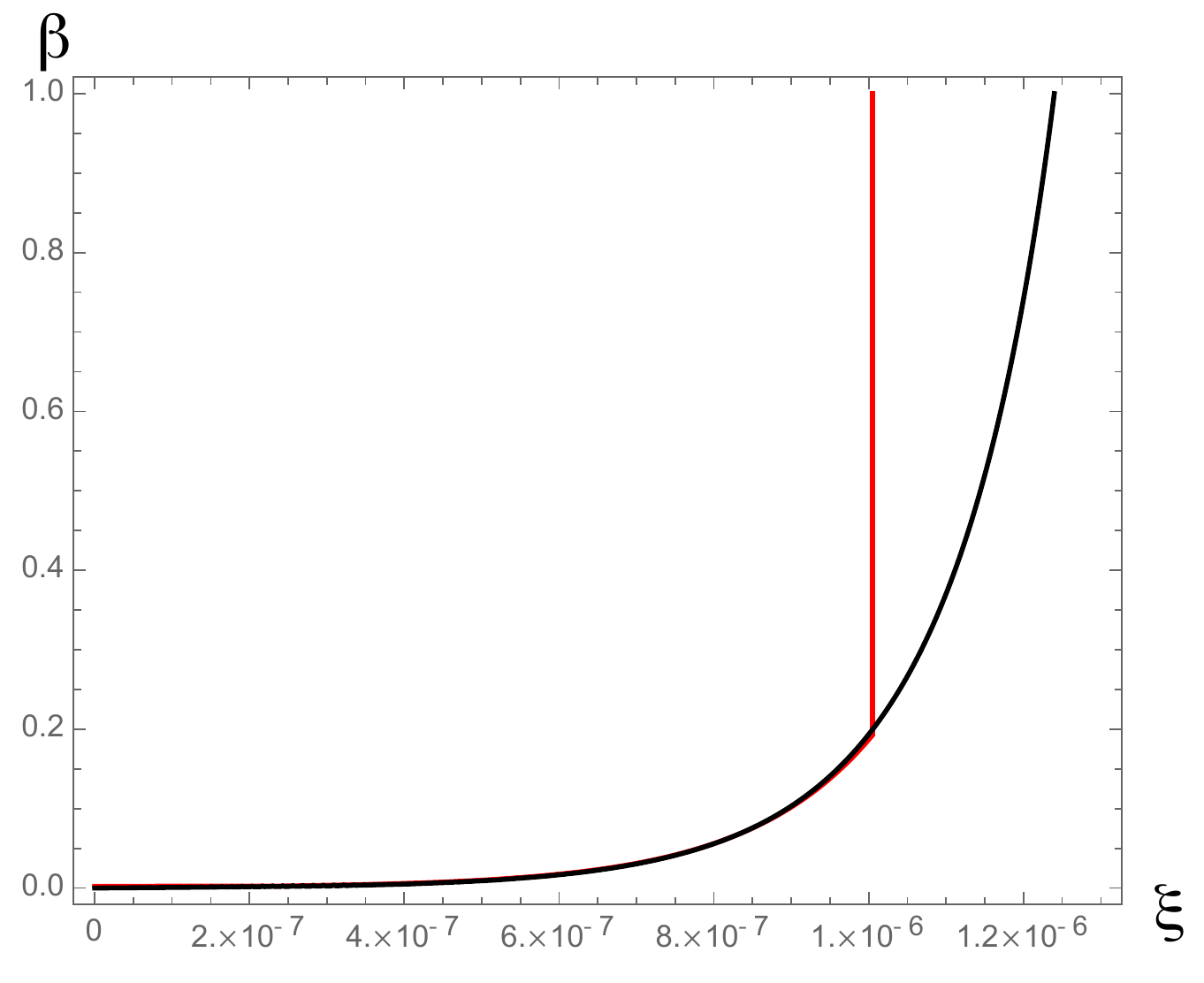}
	\caption{(Color online) Solution of Eq.~\eqref{bound3} with (i) original Read-Shockley's $f_2(\beta)=\frac{1}{4\pi } \ln (\beta_*/\beta)$ for $\beta<\beta_*$ and $0$ otherwise: red, (ii) modified $\tilde{f}_2(\beta)$ from \eqref{bound6}: black.}
	\label{BoundaryCondition}
\end{figure}

In terms of the introduced quantities the governing equations read
\begin{eqnarray}
\frac{\partial \tilde{\tau}_Y}{\partial \tilde{\omega}} = \tilde{r} - \frac{\tilde{q}(\tilde{\tau}_Y,\tilde{\rho})}{\tilde{\phi}_0}, \label{tau} 
\\
\frac{\partial \tilde{\rho}}{\partial \tilde{\omega}} = K_\rho \,\frac{\tilde{\tau}_Y\,\tilde{q}}{\tilde{\nu}(\theta,\tilde{\rho},\tilde{\phi}_0\tilde{r})^2\,\tilde{\phi}_0}\, \Bigl[1 -\frac{\tilde{\rho}}{\tilde{\rho}_{ss}(\tilde{\chi})} \Bigr],  \label{rho}
\\ 
\frac{\partial \tilde{\chi }}{\partial \tilde{\omega}} = K\,\frac{\tilde{\tau}_Y\,\tilde{q}}{\tilde{\phi }_0}\,\Bigl( 1 -\frac{\tilde{\chi}}{\tilde{\chi}_0} \Bigr), \label{chi}
\\ 
\tilde{\omega }\tilde{r}-\beta -\tilde{\tau}_B -\tilde{\tau}_Y =0, \label{force}
\end{eqnarray}
where $\tilde{\tau}_B$ is equal to
\begin{equation}
\label{backstress1}
\tilde{\tau}_B=-\frac{k_1\xi^2+(2k_0k_1-1)\xi+k_1k_0^2-2k_0}{4\pi(k_0+\xi)^2}\eta^2(\beta_{,\tilde{r}\tilde{r}}+\beta_{,\tilde{r}}/\tilde{r}-\beta/\tilde{r}^2),
\end{equation}
with $\xi=\eta |\beta_{,\tilde{r}}+\beta/\tilde{r}|$. To solve this system of partial differential equations subject to initial and boundary conditions numerically, we discretize the equations in the interval $(0,1)$ by dividing it into $n$ sub-intervals of equal length $\Delta \tilde{r}=1/n$. The first and second spatial derivative of $\beta(\tilde{r})$ in equation (\ref{backstress1}) are approximated by the finite differences
\begin{eqnarray}
\frac{\partial \beta}{\partial \tilde{r}}(\tilde{r}_i)=\frac{\beta_{i+1}-\beta_{i-1}}{2\Delta \tilde{r}},
\\
\frac{\partial ^2\beta}{\partial \tilde{r}^2}(\tilde{r}_i)=\frac{\beta_{i+1}-2\beta_i+\beta_{i-1}}{(\Delta \tilde{r})^2},
\end{eqnarray}
where $\beta_i=\beta(\tilde{r}_i)$. For the end-point $\tilde{r}=1$ we introduce $\beta_{n+1}$ at a fictitious point $\tilde{r}_{n+1}=(n+1)\Delta \tilde{r}$ and pose the discretized boundary condition
\begin{equation}
\label{bound3}
f_1(\xi_n)+f_2(\beta_n)=\tilde{\gamma}_D.
\end{equation}
with $\xi_n=\eta\frac{\beta_{n+1}-\beta_{n-1}}{2\Delta \tilde{r}}$, $\tilde{\gamma}_D=\gamma _D/(\mu b^2)$,
\begin{equation}
\label{bound4}
f_1(\xi )=-\frac{\xi}{4\pi(k_0+\xi)}-\frac{\ln (k_0+\xi)}{4\pi}+\frac{k_1\xi}{4\pi}-0.105,
\end{equation}
and
\begin{equation}
\label{bound5}
f_2(\beta)=\frac{1}{4\pi } \left( \ln \frac{\beta _*}{\beta }+\alpha \beta \right) .
\end{equation}
To avoid the singularity of $f_2(\beta)$ at $\beta =0$ which is difficult to handle numerically, we replace $f_2(\beta)$ from \eqref{bound5} by
\begin{equation}
\label{bound6}
\tilde{f}_2(\beta)=\frac{1}{4\pi } \left( \ln \frac{\beta _*}{\beta +\delta }+\alpha \beta \right) ,
\end{equation} 
with $\delta$ being a small positive number, chosen such that Eq.~\eqref{bound3} yields the curve in the $(\xi, \beta)$-plane starting from the origin (see Fig.~\ref{BoundaryCondition}). With $f_1(\xi )$ and $\tilde{f}_2(\beta)$ from Eq.~\eqref{bound4} and Eq.~\eqref{bound6}, respectively, we find that $\delta= \beta_*/(k_0 \exp(4\pi (\tilde{\gamma }_D +0.105)))$. With $\beta_{n+1}$ found from \eqref{bound3} it is possible again to discretize the first and second derivative of $\beta (\tilde{r})$ at $\tilde{r}=1$ and write the finite difference equation for $\beta (\tilde{r})$ at that point. In this way,  we reduce the four partial differential equations to a system of $4n$ ordinary differential-algebraic equations that will be solved by Matlab-ode15s. 

\begin{figure}[htp]
	\centering
	\includegraphics[width=.65\textwidth]{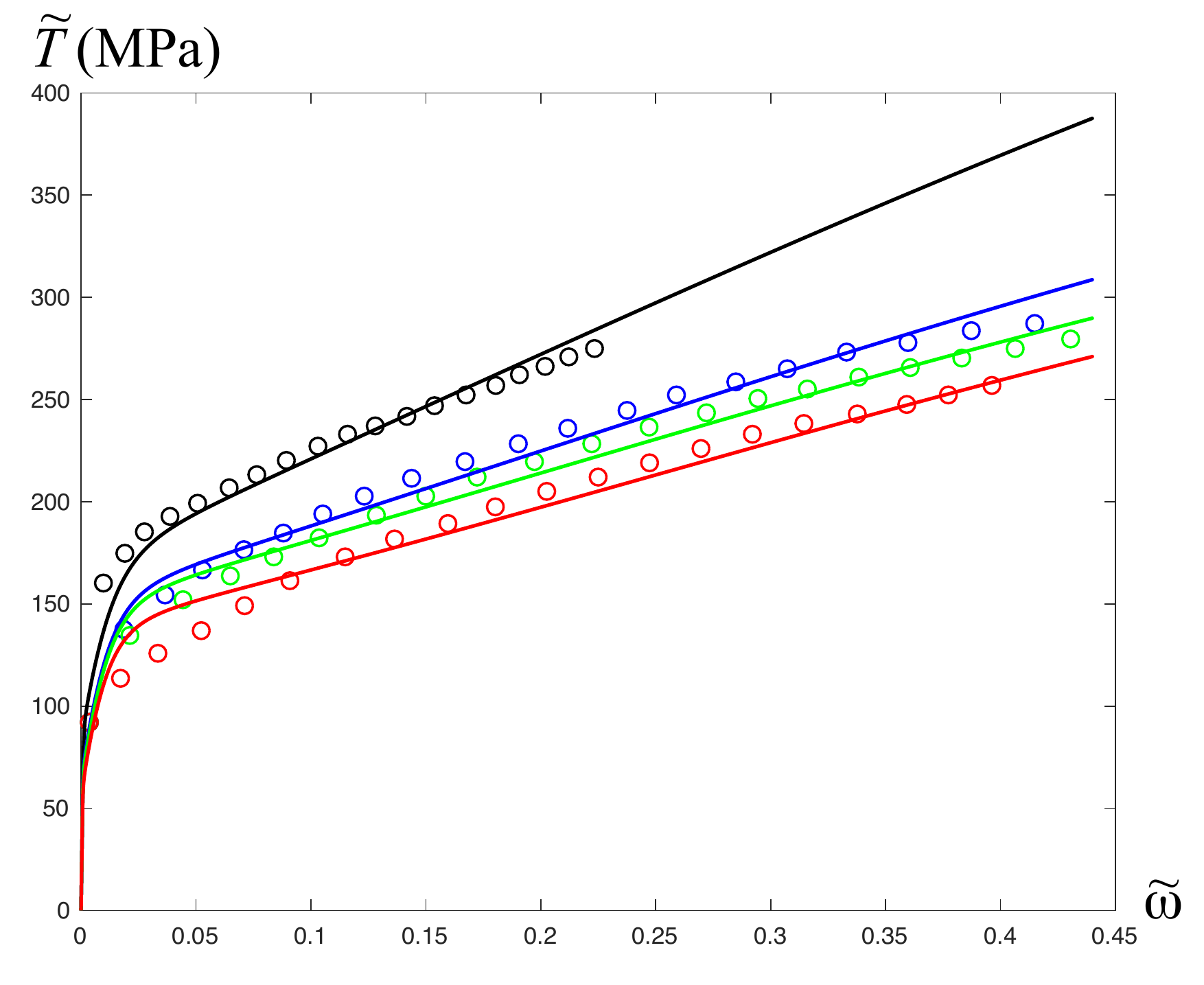}
	\caption{(Color online) Torque-twist curves with twist rate $L\dot{\omega}=\frac{\pi}{30}/$s and for room temperature, for four different radii $R=9$ micron (black), $R=15$ micron (blue), $R=21$ micron (green) and $R=52.5$ micron (red): (i) TDT (bold lines), (ii) experimental points taken from Liu {\it et al.} \cite{Liu2012} (circles).}
	\label{TorqueTwist}
\end{figure}

After finding the plastic distortion $\beta (\tilde{r})$ we compute the dimensionless torque as function of the twist angle according to
\begin{equation}
\label{torque}
\tilde{T}=T/R^3 = 2\pi \mu \int_0^1 [\tilde{\omega}\tilde{r} -\beta (\tilde{r})] \tilde{r}^2 d\tilde{r}.
\end{equation}
\begin{figure}[htp]
	\centering
	\includegraphics[width=.65\textwidth]{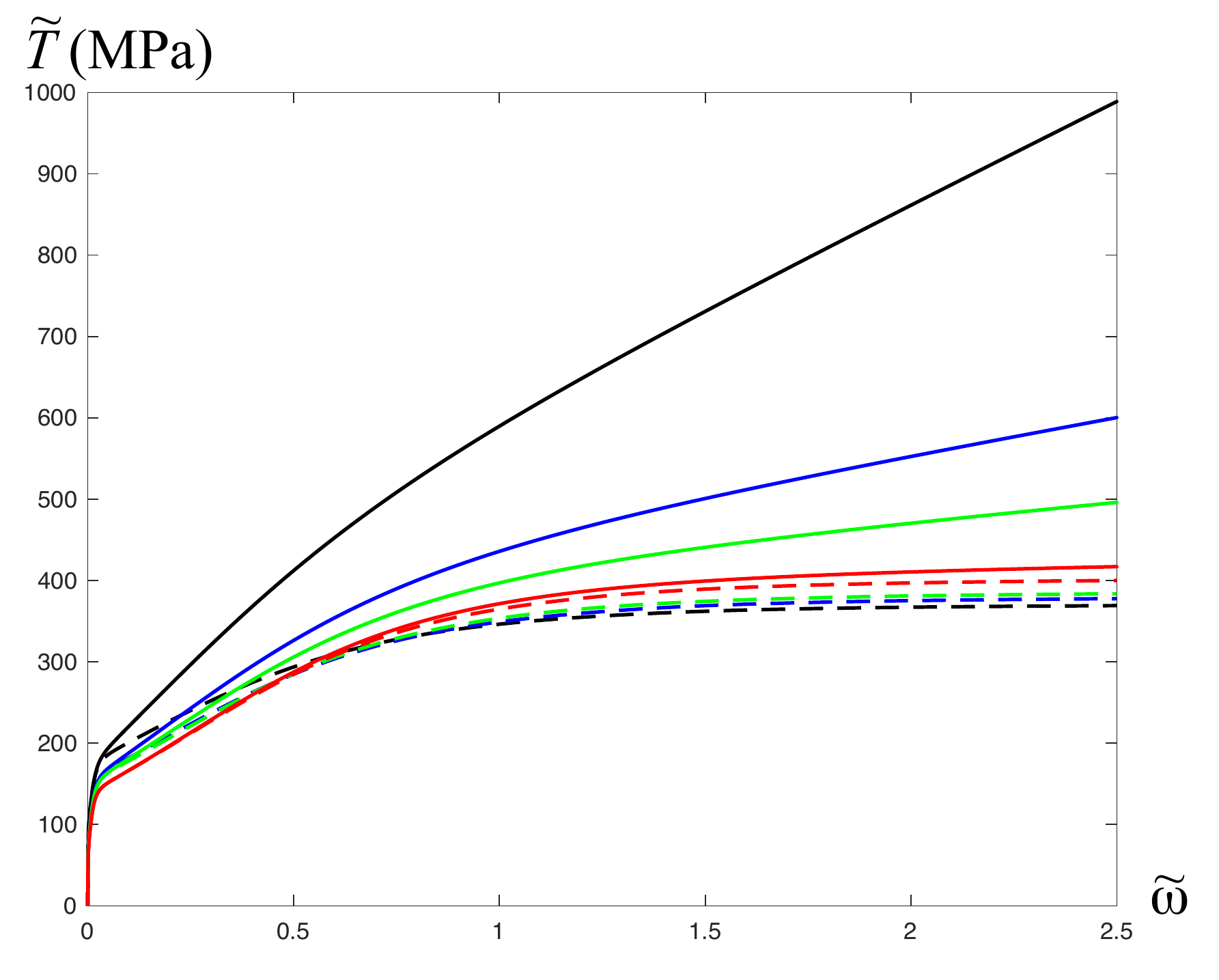}
	\caption{(Color online) Torque-twist curves with twist rate $L\dot{\omega}=\frac{\pi}{30}/$s and for room temperature, for four different radii $R=9$ micron (black), $R=15$ micron (blue), $R=21$ micron (green) and $R=52.5$ micron (red): (i) TDT (bold lines), (ii) LBL-theory (dashed lines).}
	\label{LBL-TDT}
\end{figure}

\begin{figure}[htp]
	\centering
	\includegraphics[width=.65\textwidth]{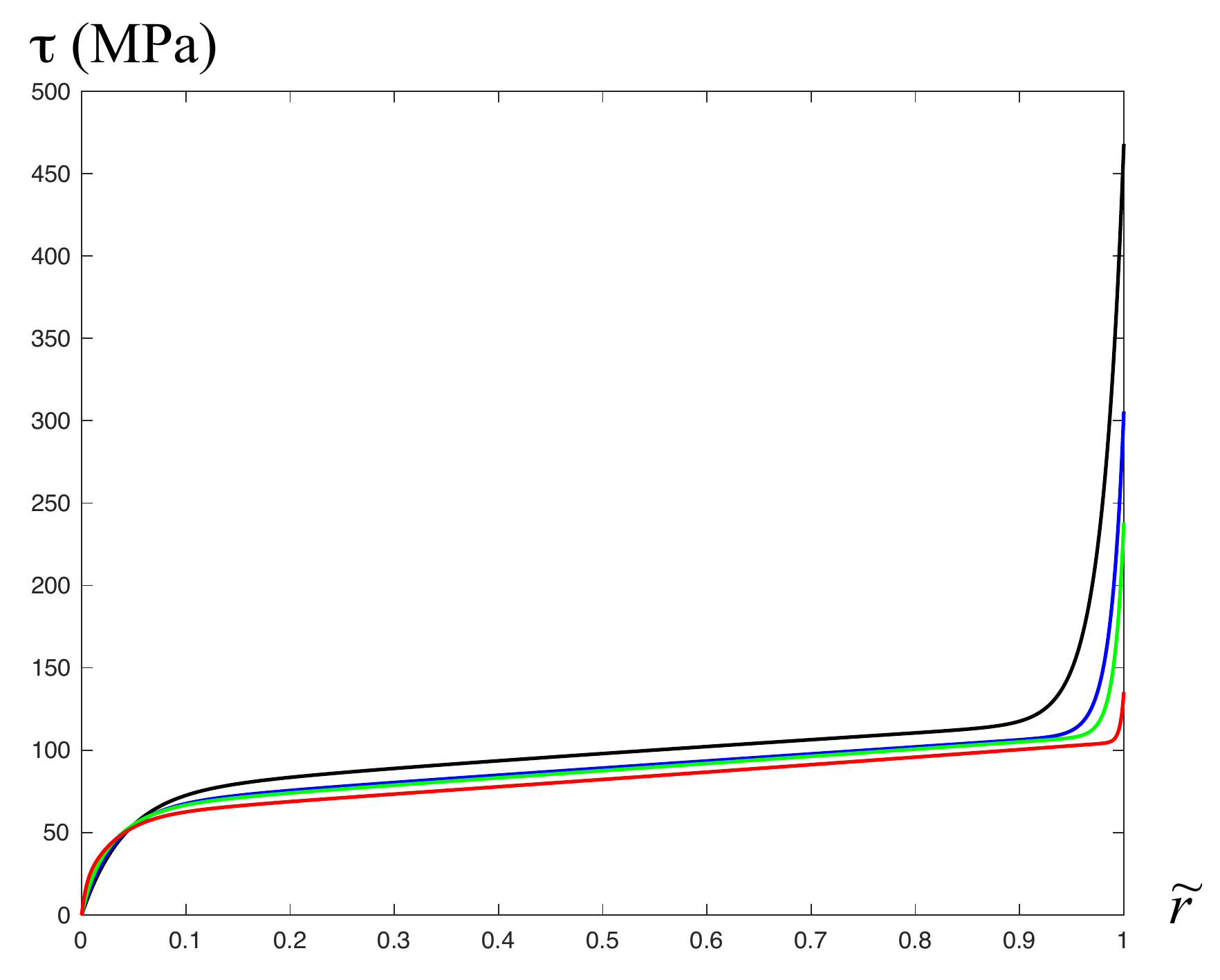}
	\caption{(Color online) Stress distributions $\tau(\tilde{r})$ at twist rate $L\dot{\omega}=\frac{\pi}{30}/$s and for room temperature, at $\tilde{\omega}=0.198$ and for four different radii $R=9$ micron (black), $R=15$ micron (blue), $R=21$ micron (green), and $R=52.5$ micron (red).}
	\label{Stress}
\end{figure}

\section{Parameter identification and numerical simulations}
\label{PI}

\begin{figure}[htp]
	\centering
	\includegraphics[width=.65\textwidth]{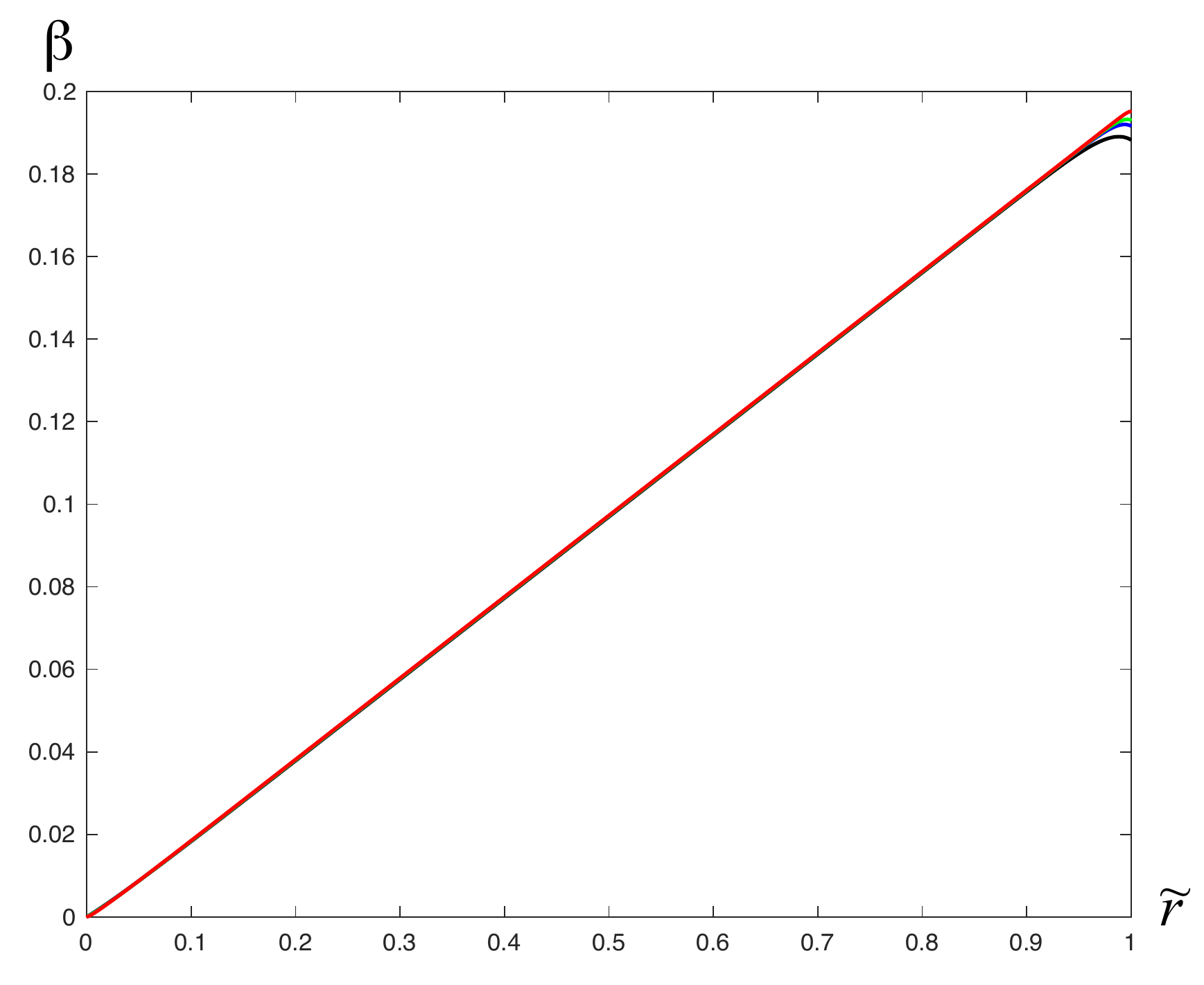}
	\caption{(Color online) Plastic distortion $\beta(\tilde{r})$ at twist rate $L\dot{\omega}=\frac{\pi}{30}/$s and for room temperature, at $\tilde{\omega}=0.198$ and for four different radii $R=9$ micron (black), $R=15$ micron (blue), $R=21$ micron (green), and $R=52.5$ micron (red).}
	\label{Beta}
\end{figure}

The experimental data of Liu {\it et al.} \cite{Liu2012} include four torque-twist curves for polycrystalline copper wires with different radii $R=9$ micron, $R=15$ micron, $R=21$ micron, and $R=52.5$ micron. Torsion tests were performed at room temperature, and for all tests the twist rate $L\dot{\omega}$ was $\pi/30$ per second ($6^\circ/$s). We show these data together with our theoretical results based on the TDT in Fig.~\ref{TorqueTwist}.  In this figure, the circles represent the experimental data in \cite{Liu2012}, while the bold lines are our theoretical simulation based on the TDT. For comparison, we show on Fig.~\ref{LBL-TDT} the torque-twist curves simulated by the TDT (bold lines) and the LBL-theory (dashed lines) in the larger range of the twist angle $\tilde{\omega}\in (0,2.5)$.  

\begin{figure}[htp]
	\centering
	\includegraphics[width=.65\textwidth]{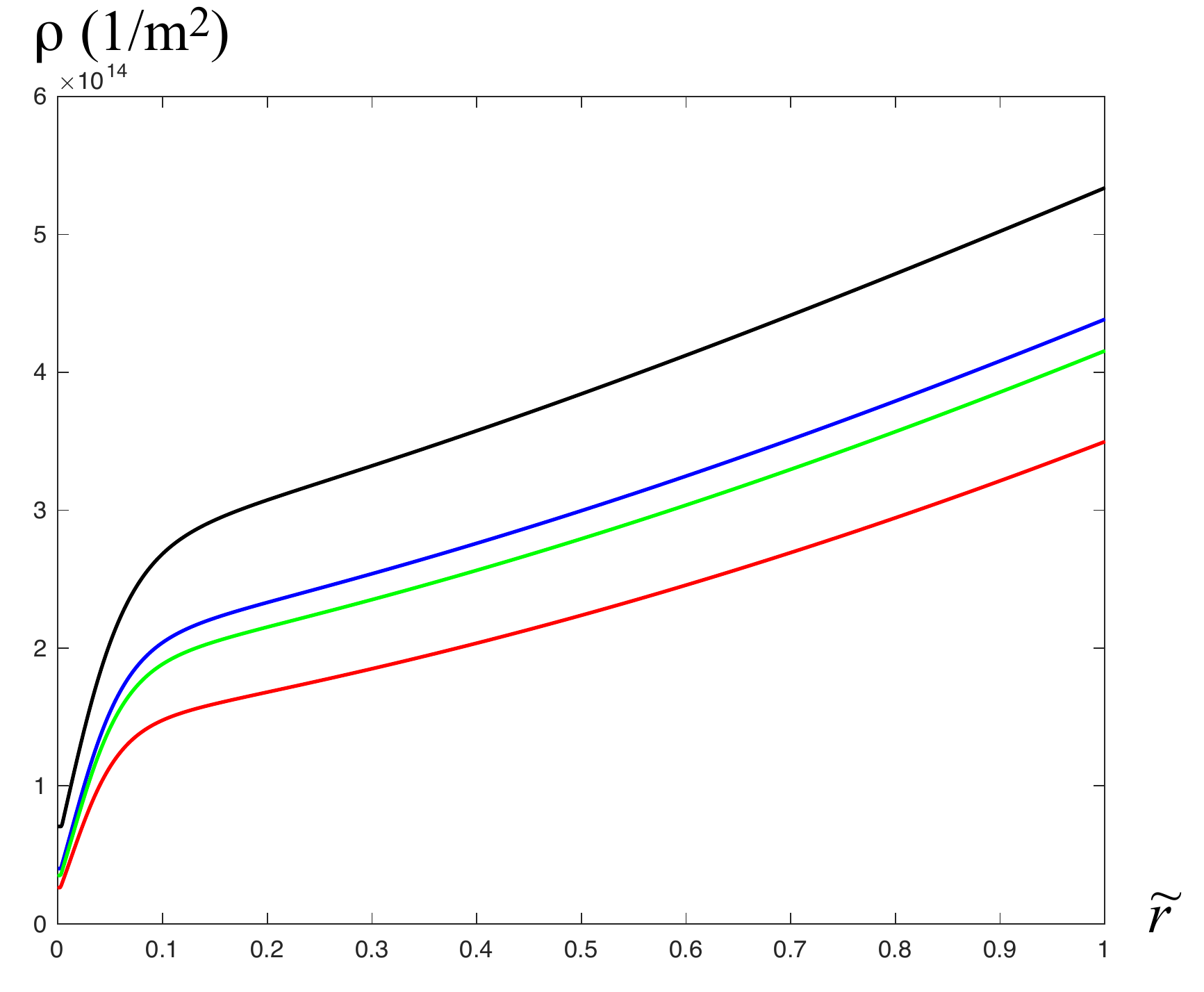}
	\caption{(Color online) Total density of dislocations $\rho(\tilde{r})$ at twist rate $L\dot{\omega}=\frac{\pi}{30}/$s and for room temperature, at $\tilde{\omega}=0.198$ and for four different radii $R=9$ micron (black), $R=15$ micron (blue), $R=21$ micron (green), and $R=52.5$ micron (red).}
	\label{DensityTotal}
\end{figure}

In order to compute the theoretical torque-twist curves, we need values for ten system-specific parameters.  The ten basic parameters are the following: the activation temperature $T_P$, the stress ratio $s$, the steady-state scaled effective temperature $\tilde\chi_0$, the two dimensionless conversion factors $K_\rho$ and $K$, the two coefficients $k_0$, and $k_1$ defining the function $\tilde{\tau}_B$ in Eq.~(\ref{backstress1}), and the three coefficients $\beta_*$, $\alpha $, and $\tilde{\gamma}_D$ entering Eq.~\eqref{bound3}. We also need initial values of the scaled dislocation density $\tilde\rho_i$ and the scaled effective disorder temperature $\tilde\chi_i$; all of which are determined by the sample preparation and microstructure of the material. For example, the history of metal forming and heat treatment may affect grain size and $\tilde\rho_i$ and $\tilde\chi_i$, which may vary from sample to sample. Note that the size of the sample can also play a role in this history of preparation. Thus, for four samples we will need to identify eight initial values. The other parameters required for numerical simulations but known from the experiment are: the ambient temperature $T=298$K, the shear modulus $\mu=48$GPa, the length $L=25$mm of the wires, the magnitude of Burgers' vector $b=2.55$\AA, the twist rate $L\dot{\omega}=\pi/30/$s, and consequently, $\phi_0=0.419\times 10^{-12}/$m. We take $a=10b$.

\begin{figure}[htp]
	\centering
	\includegraphics[width=.65\textwidth]{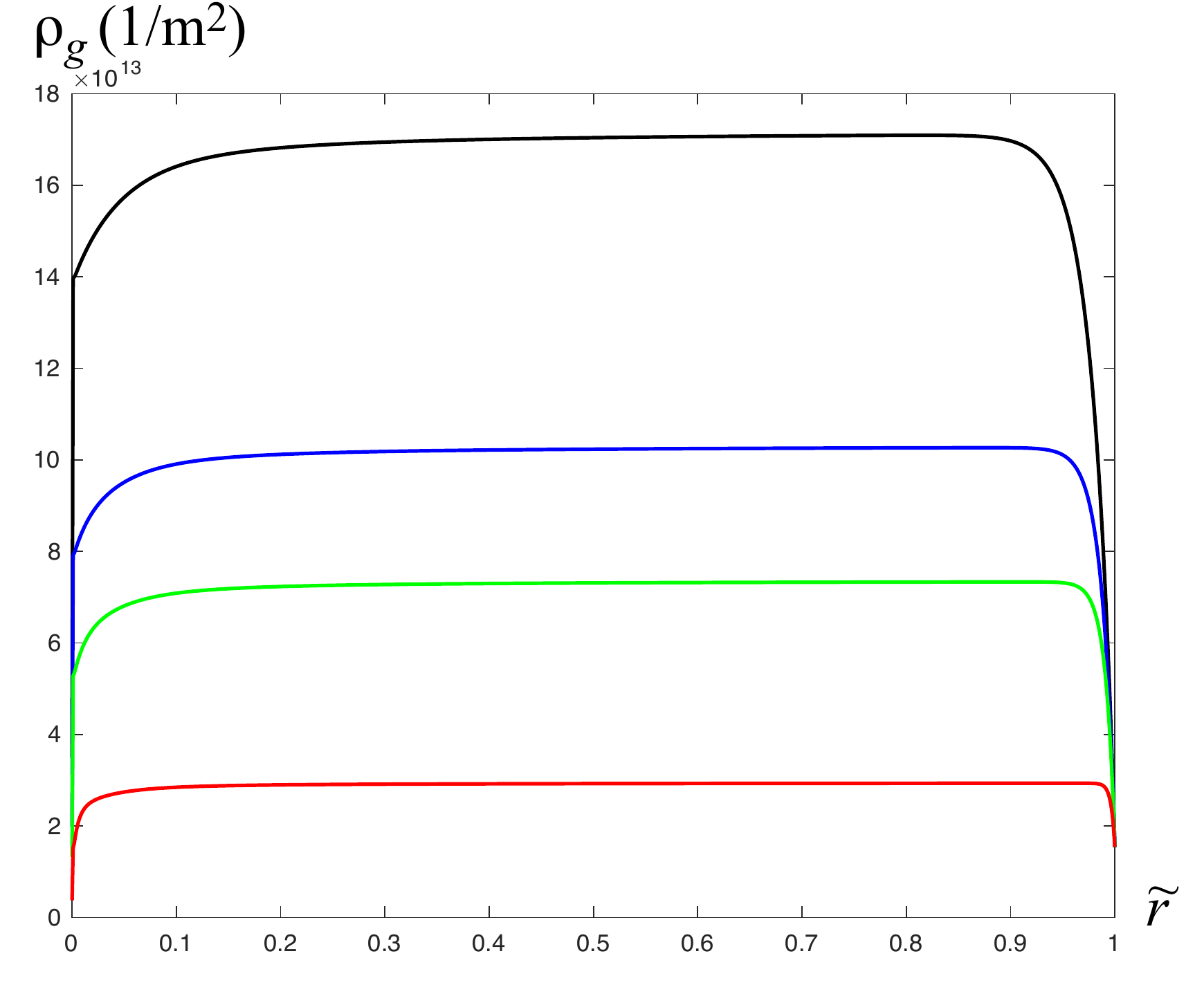}
	\caption{(Color online) Density of excess dislocations $\rho_g(\tilde{r})$ at twist rate $L\dot{\omega}=\frac{\pi}{30}/$s and for room temperature, at $\tilde{\omega}=0.198$ and for four different radii $R=9$ micron (black), $R=15$ micron (blue), $R=21$ micron (green), and $R=52.5$ micron (red).}
	\label{DensityExcess}
\end{figure}
 
In earlier papers dealing with the uniform deformations \cite{LBL-10}-\cite{JSL-17}, it was possible to begin evaluating the parameters by observing steady-state stresses $\sigma_{ss}$ at just a few strain rates $q_0$ and ambient temperatures $T_0 = T_P\,\tilde\theta_0$. Knowing $\sigma_{ss}$, $T_0$ and $q_0$ for three stress-strain curves, one could solve equation 
\begin{equation}
\label{qdef2}
\sigma = \sigma_T(\tilde\rho)\,\nu(\tilde\theta,\tilde\rho,q_0),
\end{equation} 
which is the inverse of Eq.~(\ref{qdef}) for $T_P$, $s$, and $\tilde\chi_0$, and check for consistency by looking at other steady-state situations. With that information, it was relatively easy to evaluate $K_\rho$ and $K$ by directly fitting the full stress-strain curves.  This strategy does not work here because the stress state of twisted bars is non-uniform. We may still have local steady-state stresses as function of the radius $r$, but it is impossible to extract this information from the experimental torque-twist curve. Furthermore, the similar parameters for copper found in \cite{LBL-10}-\cite{JSL-17} cannot be used here, since we are dealing with torsional deformations having the energy barrier $T_P$ and other characteristics different from those identified in the above references.

\begin{table}
  \centering 
\begin{tabular}{|c|c|c|c|c|}
\hline
$R$\,($\mu$m) & 9 & 15 & 21 & 52.5 \\ \hline
$\tilde{\rho}_i$ & $4.589\times 10^{-4}$ & $2.605\times 10^{-4}$ & $2.282\times 10^{-4}$ & $1.707\times 10^{-4}$ \\  
\hline
$\tilde{\chi}_i$ & 0.158 & 0.151 & 0.149 & 0.143 \\ \hline
\end{tabular}  
  \caption{The initial values of $\tilde{\rho}_i$ and $\tilde{\chi}_i$}\label{table:1}
\end{table}

To counter these difficulties, we have resorted to the large-scale least-squares analyses that we have used in \cite{Le17,Le18,LeTr17}. That is, we have solved the discretized system of ordinary differential-algebraic equations (DAE) numerically, provided a set of material parameters and initial values is known. Based on this numerical solution we then computed the sum of the squares of the differences between our theoretical torque-twist curves and a large set of selected experimental points, and minimized this sum in the space of the unknown parameters. The DAEs were solved numerically using the Matlab-ode15s, while the finding of least squares was realized with the Matlab-globalsearch. To keep the calculation time manageable and simultaneously ensure the accuracy, we have chosen $n=1000$ and the $\tilde{\omega}$-step equal to $0.44\times 10^{-4}$. We have found that the torque-twist curves taken from \cite{Liu2012} can be fit with just a single set of system-specific parameters.  These are: $T_P = 19205$\,K,\,$s = 0.0686,\,\chi_0= 0.2089,\,K_\rho=57.02,\, K=242.8,\,k_0 = 6.386 \times 10^{-7},\,k_1= 6.947\times 10^{6},\, \beta_*=0.192,\, \alpha =0.198,\, \tilde{\gamma}_D=1.462$. The identified initial values for four samples are shown in Table 1. To obtain the actual initial dislocation densities, we must divide these values by $a^2$, resulting in the order of $10^{13}$ dislocations per square meter. We observe that the initial dislocation density and the disorder temperature decrease with increasing radius. The agreement between theory and experiment seems to us to be well within the bounds of experimental uncertainties.  Even the initial yielding transition appears to be described accurately by this theory. There are only two visible discrepancies: (i) near the yielding transition, the torques are slightly above those predicted by theory for $R=9$ micron and $R=15$ micron, (ii) at large twist angles, the torques are slightly below those predicted by theory for $R=21$ micron and $R=52.5$ micron. Nothing about this result leads us to believe that there are relevant physical ingredients missing in the theory. The comparison also shows that the LBL-theory fails in predicting the torque twist curves and the size effect, except in the vicinity of the yielding transition where the density of excess dislocations is negligibly small. Indeed, looking at the torque-twist curves predicted by the LBL-theory in Fig.~\ref{LBL-TDT} we see that the differences in torque due to the different initial dislocation density and disorder temperature do not increase as the twist angle increases. Besides, all torque-twist curves approach the steady state at large twist angles. This contradicts the behavior of the experimental torque-twist curves obtained in \cite{Fleck94,Liu2012}.

\begin{figure}[htp]
	\centering
	\includegraphics[width=.65\textwidth]{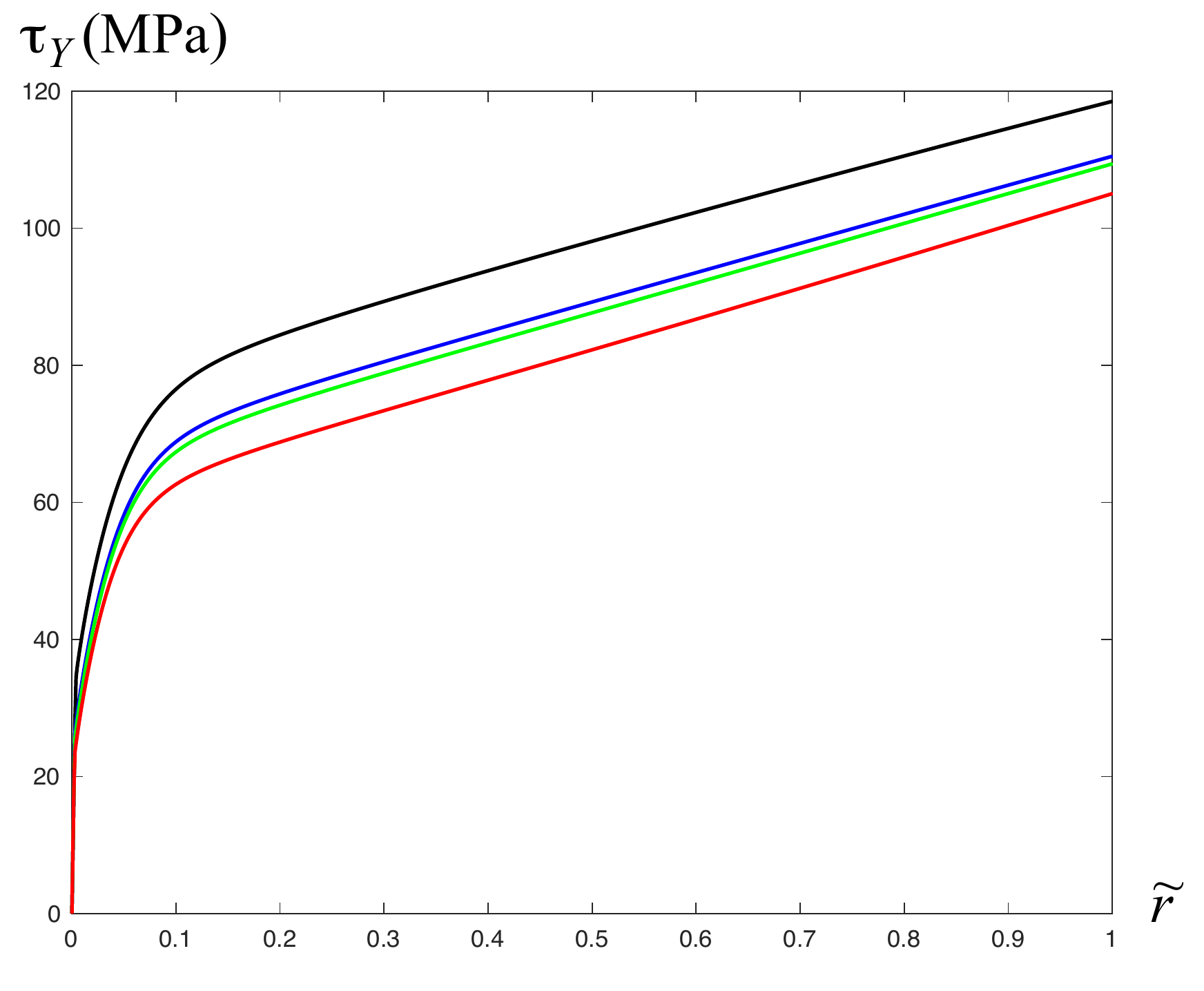}
	\caption{(Color online) Distributions of flow stress $\tau _Y(\tilde{r})$ at twist rate $L\dot{\omega}=\frac{\pi}{30}/$s and for room temperature, at $\tilde{\omega}=0.198$ and for four different radii $R=9$ micron (black), $R=15$ micron (blue), $R=21$ micron (green), and $R=52.5$ micron (red).}
	\label{FlowStress}
\end{figure}
 
\begin{figure}[htp]
	\centering
	\includegraphics[width=.65\textwidth]{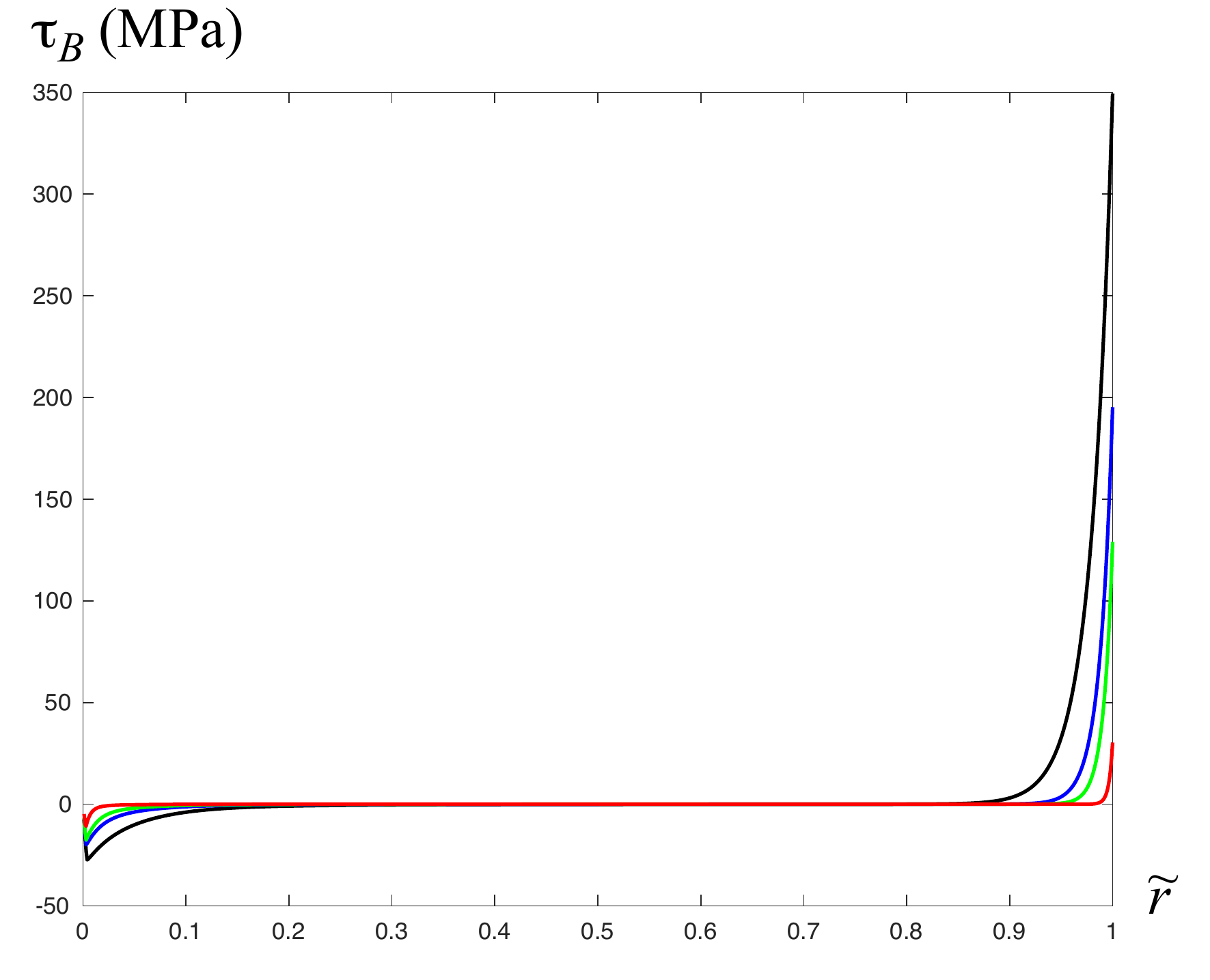}
	\caption{(Color online) Distributions of back stress $\tau _B(\tilde{r})$ at twist rate $L\dot{\omega}=\frac{\pi}{30}/$s and for room temperature, at $\tilde{\omega}=0.198$ and for four different radii $R=9$ micron (black), $R=15$ micron (blue), $R=21$ micron (green), and $R=52.5$ micron (red).}
	\label{BackStress}
\end{figure}

\begin{figure}[htp]
	\centering
	\includegraphics[width=.65\textwidth]{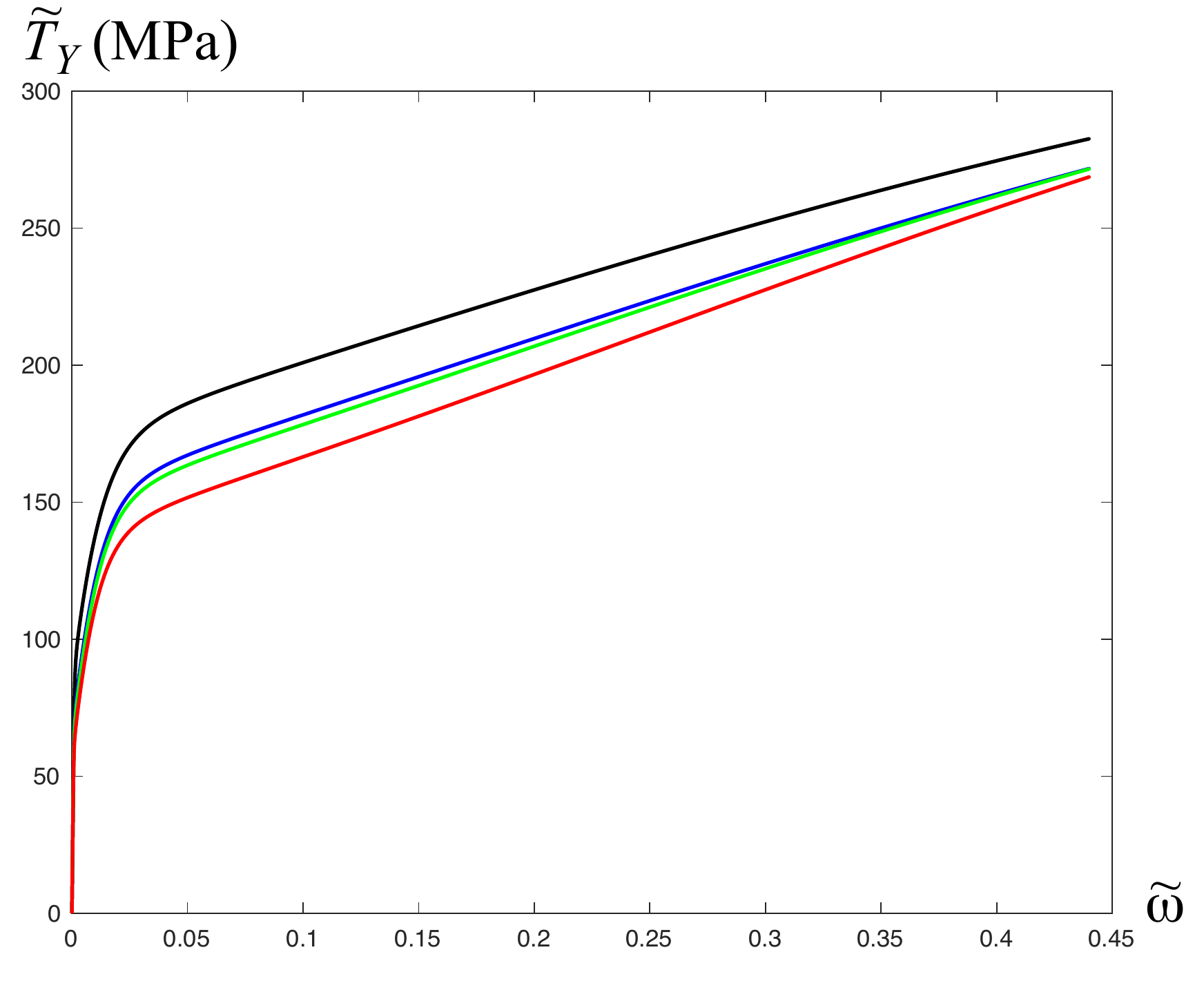}
	\caption{(Color online) Contribution of flow stress $\tau _Y(\tilde{r})$ to the torque versus twist angle at twist rate $L\dot{\omega}=\frac{\pi}{30}/$s and for room temperature, at four different radii $R=9$ micron (black), $R=15$ micron (blue), $R=21$ micron (green), and $R=52.5$ micron (red).}
	\label{FlowTorque}
\end{figure}

\begin{figure}[htp]
	\centering
	\includegraphics[width=.65\textwidth]{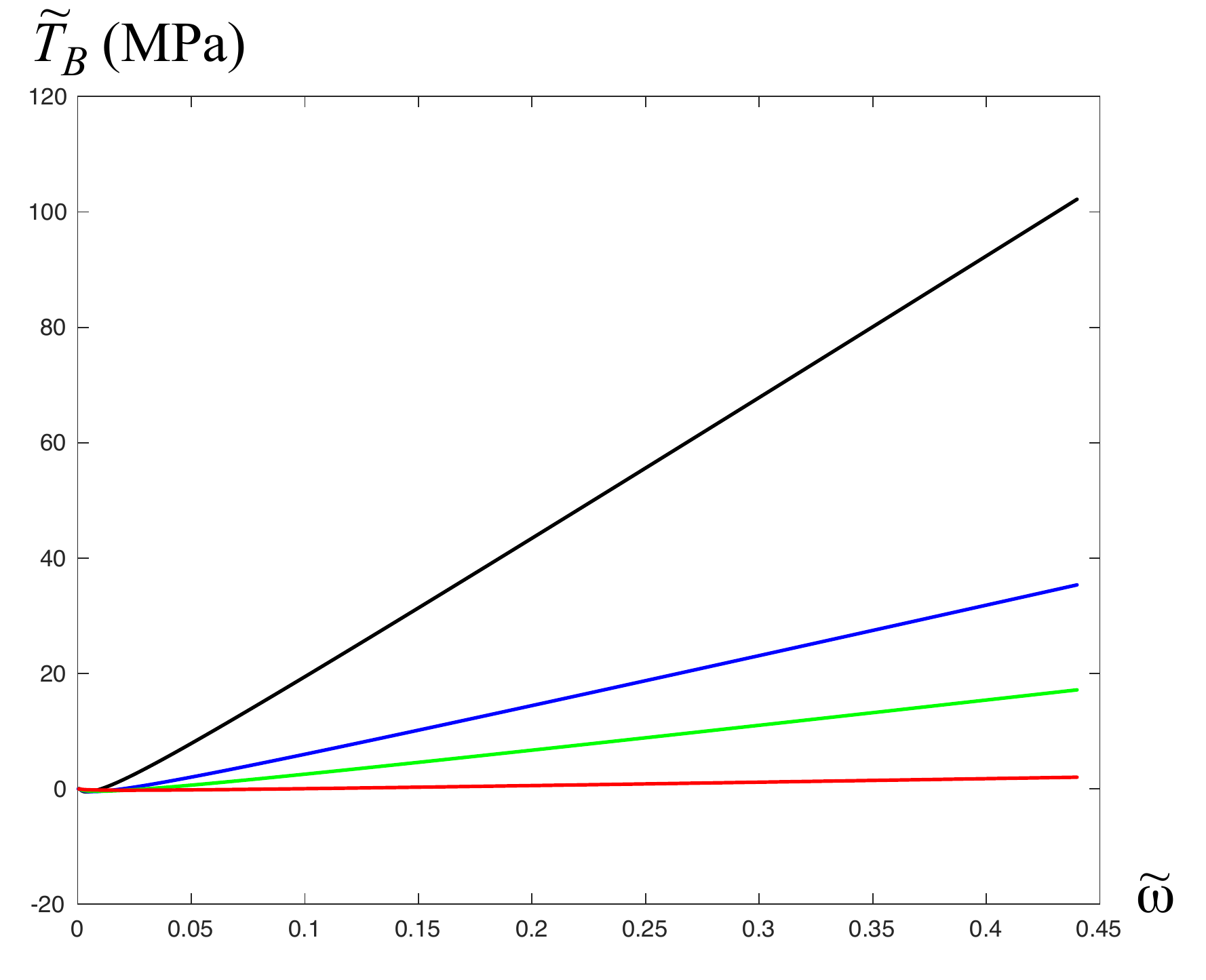}
	\caption{(Color online) Contribution of back stress $\tau _B(\tilde{r})$ to the torque versus twist angle at twist rate $L\dot{\omega}=\frac{\pi}{30}/$s and for room temperature, at four different radii $R=9$ micron (black), $R=15$ micron (blue), $R=21$ micron (green), and $R=52.5$ micron (red).}
	\label{BackTorque}
\end{figure}

The results of numerical simulations for other quantities are shown in Figs.~\ref{Stress}-\ref{BackTorque}. We plot in Fig.~\ref{Stress} the shear stress distributions $\tau (\tilde{r})=\mu (\tilde{\omega} \tilde{r} -\beta )$ at $\tilde{\omega}=0.198$. Contrary to the similar distribution obtained by the phenomenological theory of ideal plasticity, the stress in the plastic zone does not remain constant, but rises with increasing $\tilde{r}$ and reaches a maximum at $\tilde{r}=1$. The slope of this stress distribution is small in the middle ring of the cross section but increases rapidly near the free boundary. Another interesting behavior is that the stress near the center is highest for the largest wire, while it is lowest in the periphery. This behavior can later be explained by the presence of back stress. Fig.~\ref{Beta} shows distributions of the plastic distortion $\beta (\tilde{r})$ at the above twist angle $\tilde{\omega}=0.198$. It is remarkable that all four distributions of plastic distortion are nearly linear and almost indistinguishable functions of  $\tilde{r}$ except very near the free boundary $\tilde{r}=1$. Since the latter attracts excess dislocations, $\beta (\tilde{r})$ should decrease in this region leading to the decreasing density of excess dislocations. Besides, due to the different values of the factor $\eta $ for different radii, the density of excess dislocations derivable from $\beta (\tilde{r})$ depends on the radius of the wire as will be seen later. Fig.~\ref{DensityTotal} shows the total densities of dislocations at the above twist angle $\tilde{\omega}=0.198$. Note that the total density of dislocations are highest for the wire with the smallest radius. The entanglement of dislocations and the initial dislocation density and disorder temperature play a decisive role here. The higher dislocation density leads to the stronger entanglement and the higher yield stress, which, together with the higher disorder temperature, causes a stronger dislocation multiplication. Fig.~\ref{DensityExcess} presents the density of excess dislocations at the above twist angle $\tilde{\omega}=0.198$. Under the applied shear stress excess dislocations of the positive sign move to the center of the wire and pile up against the middle ring (cf. \cite{KaluzaLe11}-\cite{Liu2018}). The highest and almost constant density of excess dislocations is achieved in the middle ring, while this density decreases rapidly as $\tilde{r}$ approaches $0$ or 1. The formation and accumulation of excess dislocations in twisted wires can be explained  as follows. Since the flow stress during the plastic deformation exceeds the Taylor stress, redundant dislocations in the form of dislocation dipoles begin to dissolve according to the kinetics of thermally activated dislocation depinning \cite{LBL-10}-\cite{JSL-17}. Under the applied shear stress, positive dislocations then move towards the center and negative dislocations towards the boundary. For the dissolved dislocation dipoles within the sample and far from the free boundary, these freely moving dislocations are soon trapped by dislocations of the opposite sign. But the dislocation dipoles near the free boundary behave differently. Now the positive dislocations move inwards and become excess dislocations, while the negative dislocations leave the sample and become image dislocations. At small angles of twist, the applied shear stress near the center is still small and cannot move dislocations. Therefore, excess dislocations occupy a small outer ring. As the angle of twist increases, the shear stress increases as well, and when it becomes large enough, it can drive these excess dislocations to the center. Thus, we can say that the dissolution of dipoles near the free boundary results in excess dislocations of positive sign. They then move to the center and pile up against the middle ring, increasing the kinematic hardening.

It is interesting to examine the influence of the size of samples and the initial values of dislocation density and disorder temperature on the flow and back stress. Figs.~\ref{FlowStress} and \ref{BackStress} show the distributions of flow and back stress, respectively, at $\tilde{\omega}=0.198$. The flow stress depends linearly on $\tilde{r}$ in a small elastic zone near the center of the wire. In the plastic zone it is largest for the smallest wire. This behavior should be explained by the entanglement of dislocations as well as the initial dislocation density and disorder temperature. As shown in Table 1, the initial dislocation density $\tilde{\rho}_i$ and the initial disorder temperature $\tilde{\chi}_i$ are the highest for the smallest wire. Thus, the Taylor stress for the smallest wire is highest, and the entanglement of dislocations leads to the highest flow stress for the smallest wire. The back stress vanishes in the middle ring of the cross section because the density of excess dislocations is nearly constant there, and changes rapidly near the center and the periphery as seen in Fig.~\ref{BackStress}. The distribution of back stress shows the influence of the sample size. Since nearly the same number of excess dislocations is distributed in the wires, that wire with the smallest cross section area must have the largest magnitude of the back stress. Note however that, as the excess dislocations pile up against the middle ring, the back stress is positive in the periphery and negative near the center of the wires. This leads to the different behavior of the stress there, as can be seen in Fig.~\ref{Stress}. Figs.~\ref{FlowTorque} and \ref{BackTorque} show the contribution of the flow stress and back stress to the torque versus the twist angle. While the contribution of yield stress to torque is controlled by the entanglement of dislocations, which depends on the initial dislocation density and disorder temperature (in short, on the microstructure of the material), the contribution of back stress is controlled by the accumulation and pile-up of excess dislocations, which depend on the radius of the wire. The torque generated by the back stress from the pile-up of excess dislocations contributes a maximum of $26\%$ to the total torque in the range of the considered twist angles. Note that as the angle of twist increases, the differences in torque generated by the yield stress decrease, while the differences in torque generated by excess dislocations increase linearly.

\section{Concluding remarks}
\label{CONCLUSIONS}
  
Overall, these results seem to us to be quite satisfactory.  Note that we now use the TDT for non-uniform plastic deformations not just to test its validity but also as a tool for discovering properties of structural materials. For example, we could find the mechanism of formation and accumulation of excess dislocations based on the dissolution of dislocation dipoles near the free boundary of the bar and predict their distribution and back stress. One of the main reasons for the success of this theory -- as has been emphasized here and in earlier papers -- is the extreme sensitivity of the plastic strain rate to small changes in the temperature or the stress. Another reason for its success is the inclusion of the excess dislocations in the theory, which leads to size-dependent kinematic hardening. Here, in our opinion, the incompatible plastic distortion is the natural variable that keeps the memory of excess dislocations. It cannot enter the free energy, but the curl of this quantity should enter the free energy causing the reversible back stress. In this way the theory differs substantially from the phenomenological plasticity that introduces the back stress along with an assumed constitutive equation to fit the stress strain curves with kinematic hardening. On the contrary, our theory allows us to find the back stress from the first principle calculation of the free energy of dislocated crystals.

The results obtained show the principal applicability of TDT to non-uniform plastic deformations. We found that the behavior of the torque-twist curves is controlled not only by the microstructure of material (the grain size, the initial dislocation density, and the initial disorder temperature), which affects isotropic hardening, but above all the sample size, which affects the accumulation and pile-up of excess dislocations and kinematic hardening. For wires of micron sizes under torsion, the back stress contributes at most $26\%$ to the torque for $\tilde{\omega}<0.44$, but this contribution increases linearly with the increasing twist angle. The investigation of twisted wires under load reversal and the Bauschinger effect as well as the comparison with experiments in \cite{Liu2013} will be addressed in our forthcoming paper.

\bigskip
\noindent Acknowledgments

\smallskip
Y. Piao acknowledges financial support from the Chinese Government Scholarship Program. K.C. Le is grateful to J.S. Langer for helpful discussions and to D. Liu for informing us about the details of experimental setting in \cite{Liu2012}.

\end{document}